\documentclass[aps,prl,twocolumn,floatfix,superscriptaddress,showpacs,amsmath,amssymb]{revtex4}

\usepackage{graphicx}
\usepackage{dcolumn}
\usepackage{subfigure}
\usepackage{bm}
\usepackage{color}

\newcommand{\bmath}{\begin{displaymath}}
\newcommand{\emath}{\end{displaymath}}

\newcommand{\be}{\begin{equation}}
\newcommand{\ee}{\end{equation}}
\newcommand{\bea}{\begin{eqnarray}}
\newcommand{\eea}{\end{eqnarray}}

\newcommand{\bmultl}{\begin{multline}}
\newcommand{\emultl}{\end{multline}}

\newcommand{\bsubeq}{\begin{subequations}}
\newcommand{\esubeq}{\end{subequations}}
\newcommand{\bitemize}{\begin{itemize}}
\newcommand{\eitemize}{\end{itemize}}

\newcommand{\re}{\mathrm{Re}}
\newcommand{\im}{\mathrm{Im}}
\newcommand{\bmx}{\begin{bmatrix}}
\newcommand{\emx}{\end{bmatrix}}
\newcommand{\bsmx}{\begin{smallmatrix}}
\newcommand{\esmx}{\end{smallmatrix}}

\newcommand{\expec}[1]{\langle{#1}\rangle}

\newcommand{\Eq}[1]{Eq.~(\ref{#1})}

\newcommand{\Fig}[1]{Fig.~\ref{#1}}

\begin{document}


\title{Cavity-mediated near-critical dissipative dynamics of a driven condensate}

\author{Manas Kulkarni}
\affiliation{Department of Electrical Engineering, Princeton University,
Princeton, NJ 08544, USA}
\author{Baris \"{O}ztop}
\affiliation{Department of Electrical Engineering, Princeton University,
Princeton, NJ 08544, USA}
\affiliation{Institute of Quantum Electronics, ETH Zurich, 8093 Zurich, Switzerland}
\author{Hakan E. T\"{u}reci}
\date{\today}



\affiliation{Department of Electrical Engineering, Princeton University,
Princeton, NJ 08544, USA}
\date{\today}

\begin{abstract}
We investigate the near-critical dynamics of atomic density fluctuations in the non-equilibrium self-organization transition of an optically driven quantum gas coupled to a single mode of a cavity. In this system cavity-mediated long-range interactions between atoms, tunable by the drive strength, lead to softening of an excitation mode recently observed in experiments. This phenomenon has previously been studied within a two-mode approximation for the collective motional degrees of freedom of the atomic condensate which results in an effective open-system Dicke model. Here, including the full spectrum of atomic modes we find a finite lifetime for a roton-like mode in the Bogoliubov excitation spectrum that is strongly pump-dependent. The corresponding decay rate and critical exponents for the phase-transition are calculated explaining the non-monotonic pump-dependent atomic damping rate observed in recent experiments. We compute the near-critical behavior of the intra-cavity field fluctuations, that has been previously shown to be enhanced with respect to the equilibrium Dicke model in a two-mode approximation. We highlight the role of the finite size of the system in the suppression of it below the expectations of the open Dicke model.
\end{abstract}



\maketitle

\textit{Introduction -} 
Quantum matter coupled to enhanced optical fields in confined geometries such as resonators and waveguides offer a promising platform to study quantum dynamics and phase transitions far from equilibrium~\cite{hartmanLPR,fazio_review,hakan_nat_phys,jens}. Excitations of such systems are typically hybrid quasiparticles (polaritons) inheriting long-range coherence properties of photons and strong interactions derived from its material excitations in a controllable proportion. A crucial feature of these systems is that the fundamental light-matter interaction is particle non-conserving~\cite{marco} and the description of the system benefits immensely from an open quantum system approach. Unavoidable photon loss together with the possibility of external driving renders these systems open quantum systems which can be studied in a non-equilibrium steady state. Phase-transitions in quantum optics are not new: the laser phase transition (see~\cite{haken75} and references therein), the Dicke super-radiance transition~\cite{dicke54,lieb73-1,lieb73-2,hioe73-1,hioe73-2,walls73} and more recently weakly interacting exciton polariton condensates~\cite{iacopo} have been a corner stone of quantum optics studied extensively with the stochastic methods of open quantum systems. A new class of interacting light-matter systems attracting recent attention are systems obtained by scaling up single Cavity QED systems to lattices to simulate various Hubbard and spin models, where strong interactions of polaritonic quasi-particles lead to the possibility of studying genuine quantum many-body phenomena with photons~\cite{hartmanLPR,fazio_review,hakan_nat_phys,jens}.


While initially these systems were studied from the perspective of interacting photons i.e. quantum non-linear optics~\cite{plenio06,imamoglu09,mjh,imamoglu10}, a new aspect that has been pointed out recently in several works are mediated long-range interactions between material excitations, leading to many body phases such as a super-radiant ferro-electric phase ~\cite{marco} and spin glasses~\cite{gopalakrishnan-pra10,strack_glass} through cavity mediated long-range interactions. These two views are flip-side of the same coin and appear depending on whether photons or material excitations are integrated out in a system of mixed particles (see e.g.~\cite{marco}).

Recent work indicates that these aspects arise also for optically driven atoms coupled to cavity modes~\cite{ritsch_rmp} which were first realized experimentally for thermal atoms~\cite{vladan1,vladan2}, then for a similar system, a Bose-Einstein condensate in a fiber-based cavity \cite{colombe_nature07}, and recently with a Bose-Einstein condensate in a high-finesse optical cavity~\cite{esslinger10,esslinger_science,tilman_2013}. The latter is investigated here. In the present work, we first derive an effective Hamiltonian for the atomic subsystem by adiabatically eliminating the cavity field. We show that this results in an effective photon-mediated long-range interaction that can be expressed in terms of the photon Green's function. We then derive the full excitation spectrum of the condensate below the self-organization threshold which displays a roton-like softening at a finite momentum, as observed in recent experiments~\cite{esslinger_science}. Taking into account finite-size corrections, we show that the dynamics close to the critical point can be described by a Caldeira-Leggett model~\cite{breuer-book} with a strongly pump dependent spectral function resulting in a softening frequency for the roton-like mode.
Photon-mediated interactions lead to an effective damping of the roton-like mode that displays a non-monotonic dependence on the pump strength. Most notably, we find that the s-wave interaction is a relevant interaction that determines the nature of the dissipative dynamics near the critical point. Finally, we go beyond the adiabatic approximation and by taking into account the retarded photon-mediated interactions, we calculate the photon spectrum.


\begin{figure}[ht]
\centering
\includegraphics[scale=0.42]{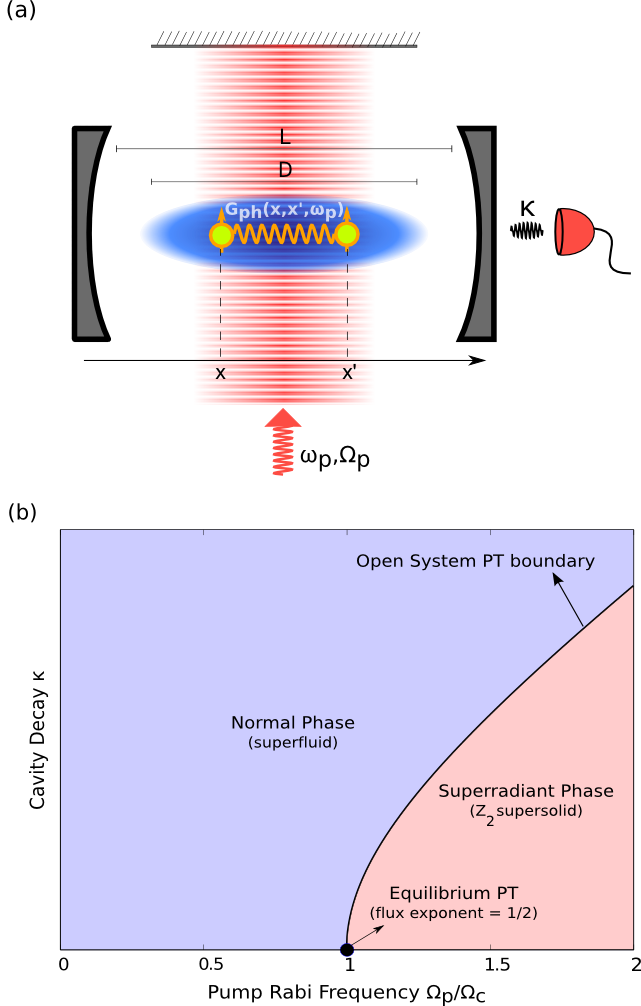}
\caption{(a)  Schematic of the setup we consider.
(b) An emerging picture showing the equilibrium phase transition point (solid black dot) and the open system phase transition boundary (black line) between the two phases.
}
\label{fig:emerging_picture}
\end{figure}

\textit{Model -} We consider a Bose-Einstein condensate (BEC) of length $D$ consisting of $N$ atoms with an internal optical transition $\omega_{a}$. This atomic condensate is coupled to a high finesse optical resonator of length $L$, mode frequencies $\omega_c^{(m)}$ and decay rates $\kappa_m$, and is driven by a laser with frequency $\omega_p$ from a direction perpendicular to the cavity axis (see~\Fig{fig:emerging_picture} (a)). We will focus here on the dispersive regime of driving ($|\Delta_a| = |\omega_p - \omega_a| \gg \gamma_{a}$, where $\gamma_{a}$ is the atomic linewidth). The Hamiltonian of the system in the rotating frame of the pump can be written as
\bea
\label{eq:mainH}
\hat{H} \!\!\! &=& \!\!\! -\hbar\sum_m \Delta_{c}^{(m)}\hat{a}_m^{\dagger}\hat{a}_m + \int dx\,\hat{\Psi}^{\dagger}(x)\left[ \frac{-\hbar^2}{2m}\frac{d^{2}}{dx^{2}} \right. \nonumber \\
&+& \hbar\frac{|\hat{E}^{+}(x)|^{2}}{\Delta_{a}}
+ \left. V_{\mbox{trap}}(x) \right] \!\! \hat{\Psi}(x) + \nonumber\\ &+&\hbar\frac{U}{2} \!\! \int \!\! dx\,\hat{\Psi}^{\dagger}(x)\hat{\Psi}^{\dagger}(x)\hat{\Psi}(x)\hat{\Psi}(x)
\eea
where $\Delta_c^{(m)} = \omega_p - \omega_c^{(m)} $ is the detuning between the pump and the $m$th mode of the cavity,  $\hat{E}^{+}(x) = g_0 \sum_m \varphi_c^{(m)} (x) \hat{a}_m + \Omega_p$ is the positive frequency component of the total electric field at point $x$ times the dipole moment of the atom. The positive rotating component of the electric field at point $x$ is composed of a contribution from the cavity $\varphi_c(x)$ and a contribution from a  coherent pump beam directed perpendicular to the cavity axis with a Rabi frequency $\Omega_p$, whose transverse profile is assumed to be uniform across the condensate (typically it's a gaussian with large beam waist). Here $g_0 \propto L^{-1/2}$ is the atom-cavity coupling strength and $\varphi_c^{(m)} (x)$ is the $m$th cavity mode function. $U$ is the strength of the s-wave interaction between the atoms. We assume $D \ll L$ as for the experimental conditions~\cite{esslinger10,esslinger_science}, imposed by the external trapping potential $V_{\mbox{trap}}(x)$ and expand the field operator as $\hat{\Psi}(x) = \sum_n \hat{c}_n \phi_n (x)$
where $\phi_n (x)$ are the eigenstates of the single particle Hamiltonian
with corresponding confinement energies $\hbar\omega_n$. For a high-finesse cavity, the above model was shown to reduce to the open Dicke Model ~\cite{nagy10, esslinger10} by keeping one cavity mode $\varphi_c (x) = \sin Gx$ at the cavity frequency $\omega_G$ closest to the pump frequency $\omega_p$ and two atomic modes, the lowest energy mode $\phi_0 (x) = 1/\sqrt{D}$ of the trap (of uniform density for our choice of the confinement potential \footnote{We model $V_{trap}$ by imposing Neumann boundary conditions at $x = (L \pm D)/2$ so that $\omega_n = \hbar\pi^2 n^2 /(2mD^2 )$ with $n = 0,1,2,\ldots$. The choice of boundary condition is for convenience and our results depend only weekly on the choice of the confinement potential in the large system size limit.}), and the mode $\phi_G (x)$ with momentum being closest to cavity wave vector $G$. As a function of the pump Rabi frequency $\Omega_p$, this model displays a phase transition from a normal phase ($\langle c_0 c_G^\dagger \rangle = \langle a_G \rangle = 0$) to a super-radiant phase ($\langle c_0 c_G^\dagger \rangle \neq 0$, $\langle a_G \rangle \neq 0$) with an atomic density modulation, observed in experiments \cite{esslinger10}.

\begin{figure}[ht]
\centering
\includegraphics[width=8.5cm]{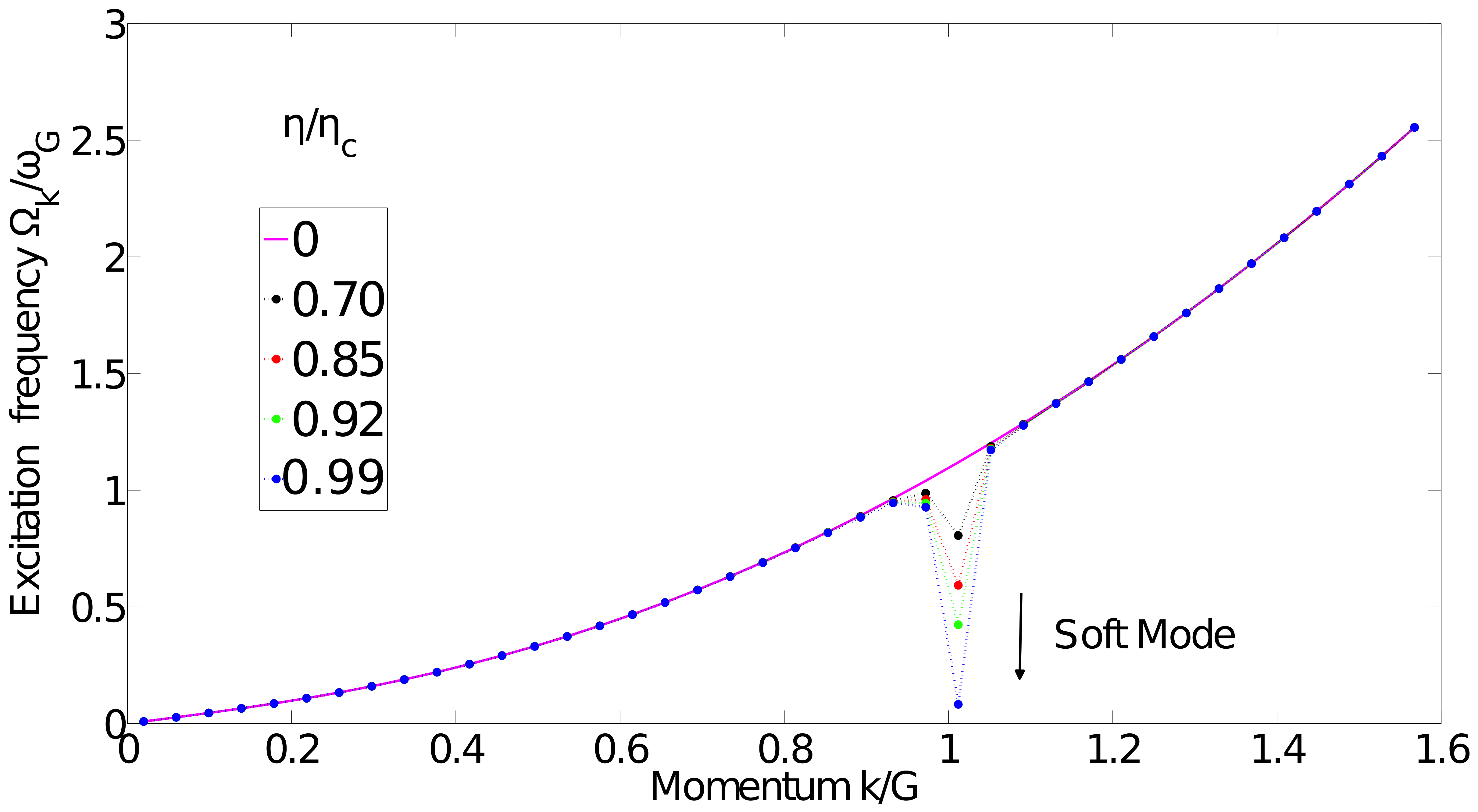}
\caption{The dispersion relation of the atomic modes for a finite value of pump is shown by dots. The solid line is the standard Bogoliubov spectrum for a BEC, at vanishing pump strength. The roton-like mode ultimately goes soft at the critical value of pump. The nearby modes are distorted (compared to the unpumped case) which we show to be a finite-size effect.
}
\label{fig:roton_new}
\end{figure}

The open nature of the system due to the leakage of cavity photons leads to a dissipative dynamics of the collective excitations of the system. A linearized fluctuation analysis~\cite{carmichael07,nagy10,oztop2012} shows that in the two-mode model, the frequency of one of the collective light-matter excitations goes to zero at the critical pump strength. Such a roton-type mode softening was observed in recent experiments~\cite{esslinger_science} through a variant of Bragg spectroscopy on a single finite-momentum mode.
The signature of this transition on various observables of the photons leaking out of the cavity was analyzed in~\cite{oztop2012} which in particular revealed that the output photon flux diverges at the critical pump strength with an exponent that is different than that of the equilibrium counterpart of the Dicke model~\cite{nagy11,oztop2012}. This observation was attributed later~\cite{torre2013} to an effective temperature \cite{gambassi1,gambassi2} generated for the low-energy degrees of freedom due to coupling of the cavity mode to vacuum fluctuations. The emerging picture is discussed in~\Fig{fig:emerging_picture}. The fluctuations that trigger the instability are atomic density fluctuations generated by long-range photon-mediated interactions. The effect of vacuum fluctuations coupled to the cavity mode is transferred to the atoms via the atom-light coupling and generates an effective low-energy temperature in the steady state. The phenomenology of the open Dicke model is then similar to that of the finite temperature Dicke model in so far as the low-energy degrees of freedom are concerned~\cite{torre2013}.

Recent experiments~\cite{tilman_2013} monitoring the photon number fluctuations in real time however find that the scaling of the diverging photon fluctuations close to the phase transition cannot be understood by cavity dissipation alone. A careful analysis revealed that there must be a residual atomic damping of unknown origin that has a highly non-trivial dependence on the pump strength. Here we first explain the origin of this atomic dissipation departing from the two-mode picture underlying the Dicke model and taking into account the full spectrum of collective atomic excitations. 

\begin{figure}[ht]
\includegraphics[scale=.125]{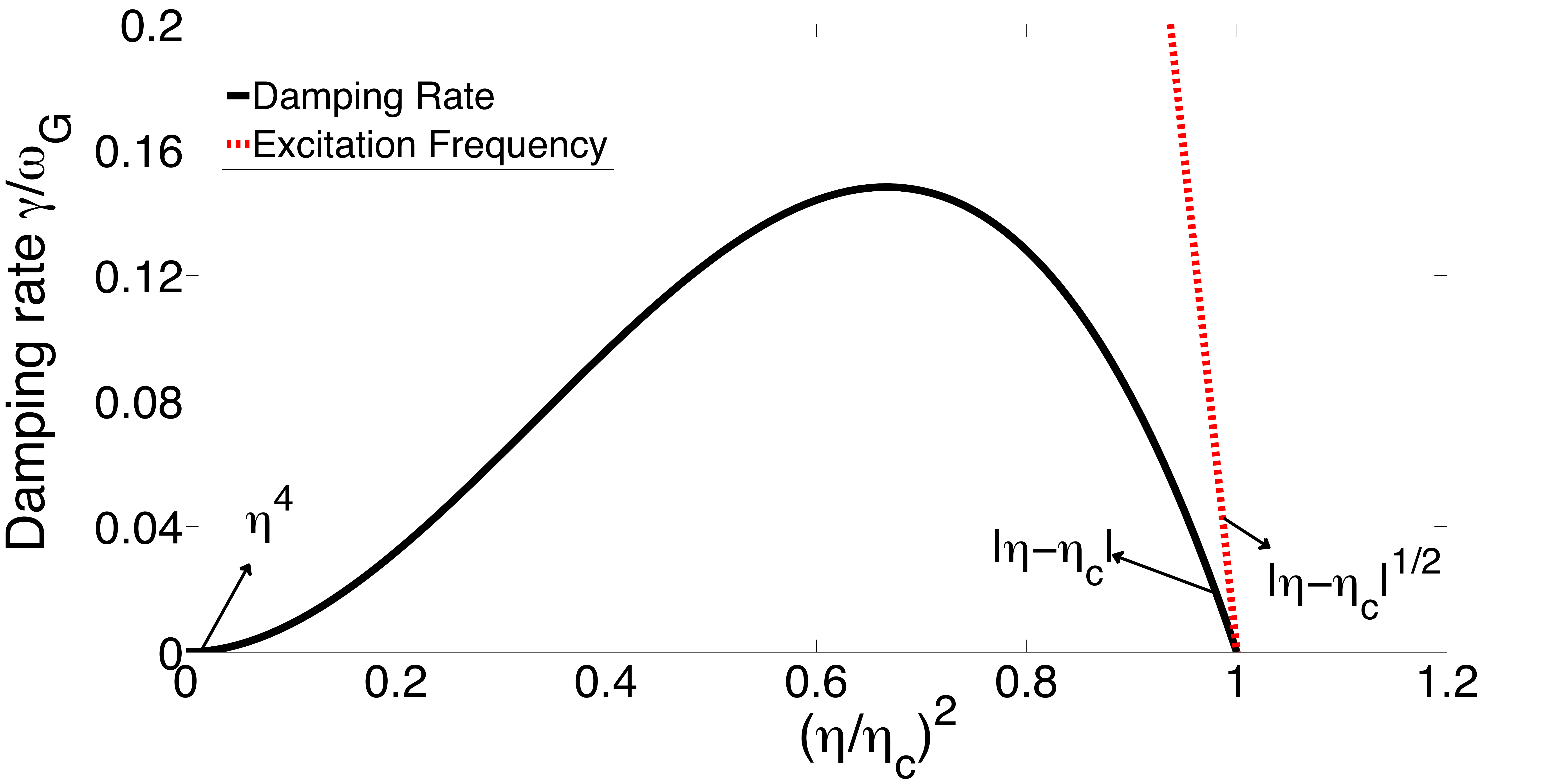}
\caption{Damping rate, $\gamma = -\im [\omega_{\mbox{rot}}]$, as a function of the pump strength is shown in black (solid line). The damping rate vanishes with a critical exponent of 1 and the excitation frequency is shown in red (dashed) which vanishes with exponent 1/2.}
\label{fig:damping_rate}
\end{figure}

\textit{Adiabatic elimination -} Under typical experimental conditions~\cite{esslinger_science}, the time scale of the cavity dynamics of the cavity modes, given by the corresponding decay rates $\kappa_m$, is much faster than that of the atomic dynamics (given the by recoil frequency $\omega_r = \hbar G^2 /2m$). In this regime, the cavity field operator can be adiabatically eliminated \cite{esslinger_science} and the Hamiltonian in~(\ref{eq:mainH}) becomes only atomic
with a nonlocal potential $U_{NL}(x,x') = 2 \hbar (g_0 \Omega_p /\Delta_a)^2 \,\, \re [G_{ph}(x,x';\omega_p )]$ where
\be
G_{ph}(x,x';\omega ) = \sum_m \frac{\varphi_c^{(m)} (x) \varphi_c^{(m)*} (x')}{\omega - \omega_c^{(m)} + i\kappa_m }
\label{eq:}
\ee
is the photon Green's function. The form of this Hamiltonian transparently illustrates the presence of long-range interactions between the atoms that are mediated by the photons. 
\footnote{It is reminiscent of the laser-induced interaction in BEC's~\cite{kurizki_prl02,kurizki_prl03} or gas made of atoms with large magnetic moments~\cite{santos}}. 
For the single cavity mode case, in the thermodynamic limit ($N,L,D \rightarrow \infty$ keeping density $N/D$ and the ratio $L/D$ constant), there is one mode at a finite momentum $G$ in the condensate excitation spectrum that softens first as the pump strength approaches its critical value. A two-mode description then captures the frequency of this {\it roton-like} mode~\cite{nagy10,esslinger_science,oztop2012} $\Omega_G \! = \omega_{G}^{\prime}\sqrt{1-\frac{\eta^{2}}{\eta_{c}^{2}}}$.
where $\omega_{G}^{\prime} = \sqrt{\omega_G(\omega_G + 2\frac{NU}{D})}$, $\omega_G \equiv \omega_{n_G}$, $\eta^2 = -2 \Delta_c (g_0 \Omega_p /\Delta_a)^2  / (\Delta_c^2 + \kappa^2 )$ and $T_{mn} = \int_{cav} dx \phi_n (x) \phi_m (x) \varphi_c (x)$. $\eta_c = \sqrt{\omega_G^{\prime2} /(2N \omega_G T_{0G}^2)}$ is the critical value of the pump dependent parameter $\eta$. We are rather interested in the large but finite $N,L,D$ regime, where the finite size effects take place. In this parameter regime, the roton-like mode (with momentum $\simeq G$) gets coupled to other modes and acquires a finite lifetime due to this coupling. The dynamics of this mode can be described by a Caldeira-Leggett model.
\bea
\frac{\delta\hat{H}}{\hbar} &=& \Omega_G \delta\hat{b}_G^{\dagger} \delta\hat{b}_G + \sum_{n\neq G} \Omega_n \delta\hat{b}_n^{\dagger} \delta\hat{b}_n \nonumber \\
&-& (\delta\hat{b}_G + \delta\hat{b}_G^{\dagger}) \sum_{n\neq G} g_n (\delta\hat{b}_n + \delta\hat{b}_n^{\dagger}),
\label{eq:mode_H_diag}
\eea
where the system frequency given by $\Omega_G$ goes to zero as $|\eta - \eta_c|^{\frac{1}{2}}$ and the bath dispersion is given by $\Omega_n = \sqrt{\omega_n^2 - 2\omega_n N\eta^2 \left(T_{0n}^2 - \frac{U}{D\eta^2} \right)}$, shown in Fig.~\ref{fig:roton_new}. Here $\delta\hat{b}$ represents atomic fluctuations after a Bogoliubov transformation. We note that the bath coupling $g_n \propto |\eta - \eta_c|^{-\frac{1}{4}}$ is strongly pump dependent as well. The resulting effective atomic bath coupling to the roton mode has a low-energy spectral density that is super-ohmic $J(\omega)\sim \frac{1}{D} \omega ^{3}$ \footnote{We note that this form is crucially dependent on the existence of s-wave interactions. For $U=0$, the low-energy spectral density would be sub-ohmic with $J(\omega) \propto \sqrt{\omega}$. The s-wave interaction is therefore a relevant perturbation for the physics we are describing here.}. As a result of the coupling to full spectrum of atomic modes, the characteristic frequency of the roton-mode acquires a real correction and an imaginary part, the latter yielding the lifetime of the roton-like mode that is acquired due to the finite size of the system. To the lowest order in $1/D$ we get
\bea
\label{eq:chfreq}
\omega_{\mbox{rot}} = \pm\Omega_{G} -\frac{i}{2}A \eta^{4} \Omega_{G}^{2}
\eea
where $A$ is a constant of $\mathcal{O}(1/D)$. The characteristic frequencies are shown in Fig.~\ref{fig:damping_rate} where real part and imaginary part have critical behavior given by $|\eta - \eta_c|^{\frac{1}{2}}$ and $|\eta - \eta_c|$ respectively.
The dynamics near the critical point is hence underdamped. The effective atomic dissipation rate is found to have a non-monotonic dependence on the pump $\eta$, as shown in Fig.~\ref{fig:damping_rate}. Hence the origin of the additional atomic damping that was found in recent experiments~\cite{tilman_2013} is finite-size induced photon-mediated coupling between atomic modes.

\textit{Beyond the adiabatic limit -} To access output photon observables, we start from the original Hamiltonian Eq. \ref{eq:mainH} and expand both the light and matter parts in terms of mean-field plus linear fluctuations ($\hat{\Psi}=\bar{\Psi}+\delta\hat{\Psi}, \hat{a}=\bar{a}+\delta\hat{a}$). This has two effects: the excitations are now collective light-matter excitations and the photon-mediated interactions are retarded. By linearizing~\Eq{eq:mainH} in photonic and atomic fluctuations we obtain
\bea
\label{eq:hamAP}
\frac{\delta \hat{H}}{\hbar} \!\!\! &=& \!\!\! -\Delta_{c}\delta \hat{a}^{\dagger}\delta \hat{a} \! + \! \omega_{G}^{\prime} \delta \hat{d}_{G}^{\dagger} \delta \hat{d}_{G} \! + \! \lambda_{G} \! \left( \! \delta \hat{d}_{G} \! + \! \delta \hat{d}_{G}^{\dagger} \right) \! \left( \delta \hat{a} \! + \! \delta \hat{a}^{\dagger} \right) \nonumber \\
\!\!\!&+& \!\!\! \sum_{n\neq G} \! \lambda_{n} \! \left(\delta \hat{d}_{n} \! + \! \delta \hat{d}_{n}^{\dagger}\right)\left(\delta \hat{a} \! + \! \delta \hat{a}^{\dagger}\right) + \sum_{n\neq G}\omega_{n}^{\prime}\delta \hat{d}_{n}^{\dagger}\delta \hat{d}_{n} \nonumber \\
&+& \sum_m \xi_m (\delta\hat{a}^{\dagger} \hat{r}_m + \hat{r}_m^{\dagger}\delta\hat{a} ) + \sum_m (\nu_m - \omega_p )\hat{r}_m^{\dagger}\hat{r}_m
\eea
where $\lambda_{n} = (g_0 \Omega_p / \Delta_a ) \sqrt{N} T_{on} \sqrt{\frac{\omega_{n}}{\omega_{n}^{\prime}}}$ with $\omega_{n}^{\prime} = \sqrt{\omega_{n}^{2} + 2\omega_{n}\frac{NU}{D}}$. Here, $\delta \hat{d}$ represents the atomic fluctuations (after a Bogoliubov transformation) and $\delta \hat{a}$ represents the photonic fluctuations. The operators $\hat{r}_m$ are used for the continuum reservoir of photons outside the cavity, $\nu_m$ are the corresponding reservoir mode frequencies and $\xi_m$ are the coupling to the cavity photons. We treat this coupling in the usual Markov approximation where the cavity decay rate becomes $\kappa = \pi |\xi (\omega_c )|^2 \Gamma (\omega_c )$ with $\Gamma$ being the density of reservoir states~\cite{scully-book}. We solve the coupled equations of motion in the Fourier domain by integrating out the atomic and photonic reservoir degrees of freedom. 
An analytic expression for photon flux can be obtained but is rather lengthy. In the absence of the atomic bath, the obtained out-coupled photon flux is 
\bea
\expec{\delta \hat{a}^{\dagger}\delta \hat{a}} = \frac{1}{-2\Delta_{c}\omega_{G}^{\prime}} \frac{\lambda_G^{2}}{\left[ 1 - \frac{\lambda_G^{2}}{\lambda_{c}^{2}} \right] }
\eea
where the critical point is given by $\lambda_{c} = \frac{1}{2} \sqrt{\frac{\omega_{G}^{\prime}}{-\Delta_{c}} \left( \kappa^{2} +\Delta_{c}^{2}\right) }$. This generalizes the result obtained in~\cite{oztop2012} to the case $U \neq 0$. Including the full spectrum of atomic modes results in a sub-dominant contribution of order $|\lambda_G-\lambda_c|^{-3}$ such that the critical point and critical exponent remain the same. Needless to say, the highly non-trivial pump-dependence of atomic dissipation channel is crucial to obtain this result.

\begin{figure}[htb]
\includegraphics[scale=0.18]{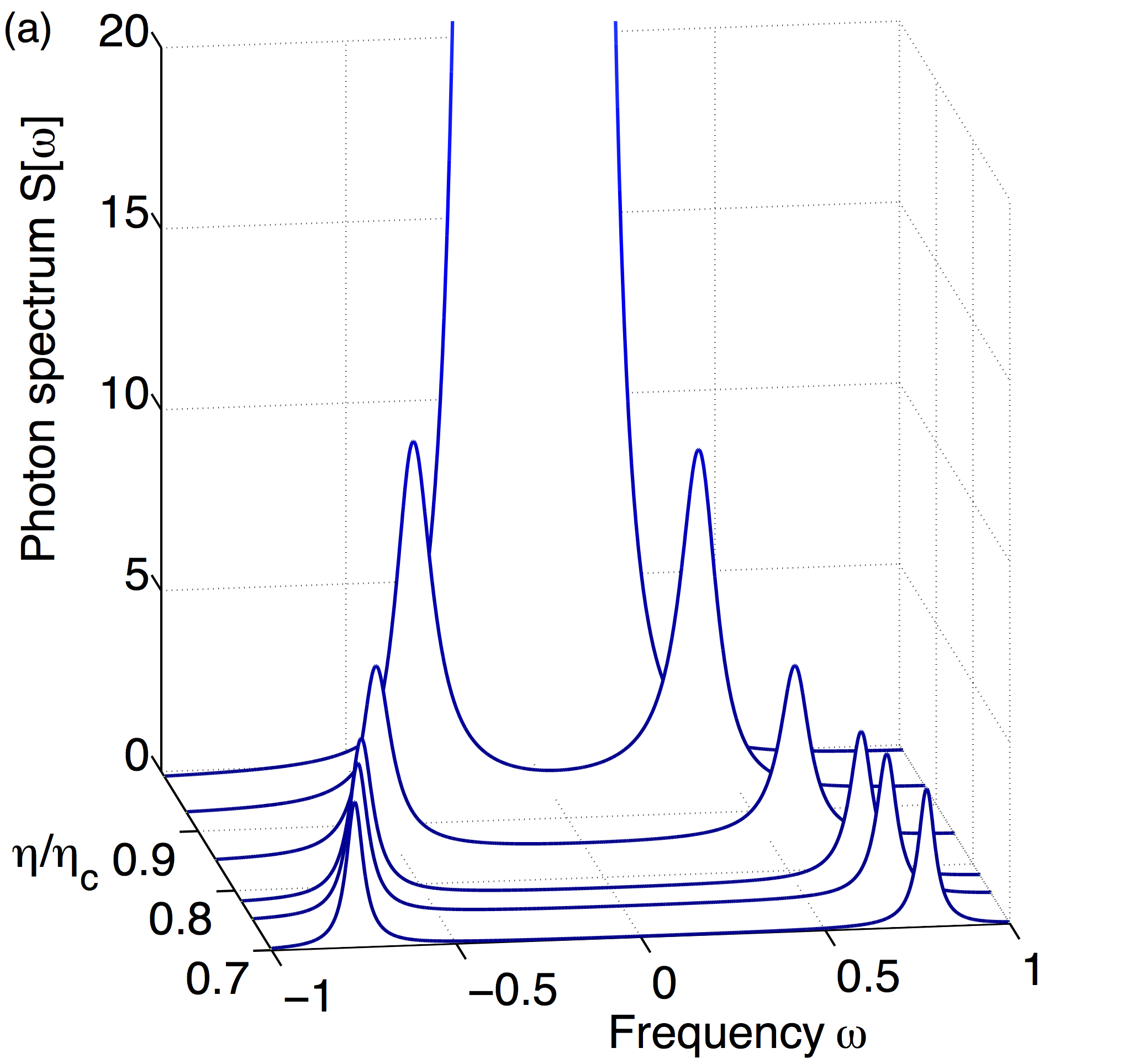} 
\includegraphics[width=8.5cm]{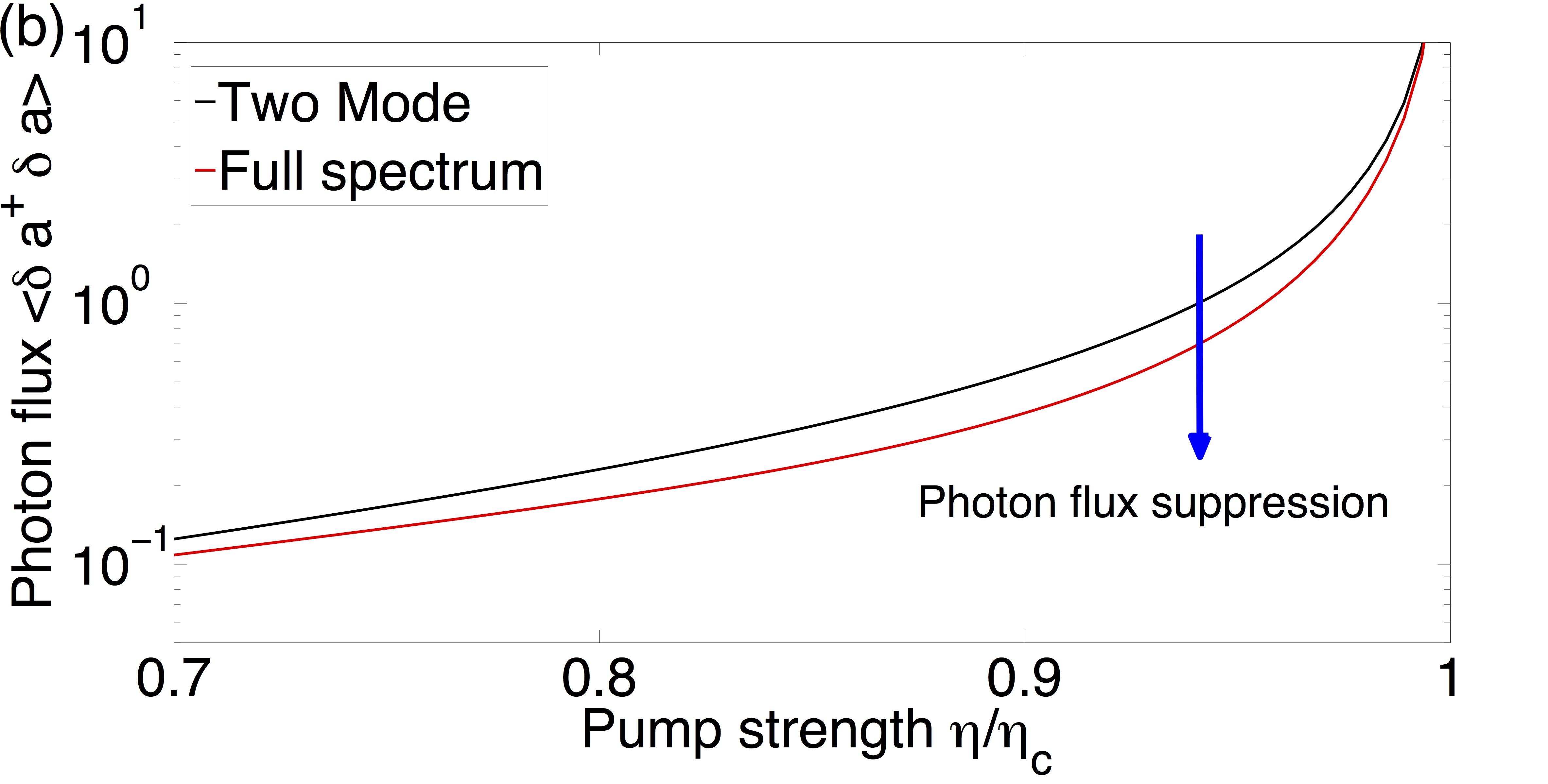}

\caption{(a) A 3D visualization of photon spectrum is shown above. It can be seen that the peaks move closer to the origin as pump increases and the peak positions move towards the origin as $|\eta - \eta_c|^{\frac{1}{2}}$ and their width becomes smaller (ie, peaks become sharper ) as $|\eta - \eta_c|$. (b) Photon flux as a function of pump in the two mode and full atomic bath picture. We find suppression due to atomic bath consistent with a recent experiment~\cite{tilman_2013}.}
\label{fig:spec3D}
\end{figure}

It is worthwhile to look at the frequency resolved spectrum of the flux shown in Fig.~\ref{fig:spec3D}, given by $S[\omega ] = \int_{-\infty}^{\infty} d\tau e^{-i\omega\tau} \expec{\delta\hat{a}^{\dagger}(\tau ) \delta\hat{a}(0)}$.
The two peaks correspond to the broadened polaritonic normal soft modes of the system, that can be described by a two-impurity Caldeira-Leggett model~\cite{caldeira_prl06,nemes_physicaa06} coupled to a common bath through a pump-dependent coupling (two other peaks corresponding to the non-soft modes are not shown here). As pump approaches the critical value, the two peaks move inwards towards zero frequency, their peak frequencies scaling as  $|\eta - \eta_c|^{\frac{1}{2}}$ and their widths as $|\eta - \eta_c|$. Very close to the critical value, the two peaks merge. At that point, the excitation frequency is purely imaginary, given to lowest order by
\bea
\label{eq:pol_freq}
\omega_{\mbox{pol}}\sim
-i\frac{\Delta_{c}^{2}+\kappa^{2}}{2\kappa}\left(1-\frac{\lambda_{G}^{2}}{\lambda_{c}^{2}}\right)
\eea
in a $\left(1-\lambda_{G}^{2}/\lambda_{c}^{2}\right)$ expansion. This is a unique feature of the inclusion of the retarded photon mediated interactions: the existence of a narrow window around the critical point of overdamped excitation dynamics. The real part vanishes at a bifurcation point for slightly lower pump values than the critical pump (this was also observed in the two-mode case in Refs.~\cite{carmichael07,nagy08,oztop2012}). It turns out that in a small $\left(1-\lambda_{G}^{2}/\lambda_{c}^{2}\right)$ expansion the atomic bath plays a role at the third order which is consistent with the observation that critical point and exponent due to the addition of atomic bath remains unchanged, although there is departure away from the critical point.
From Fig.~\ref{fig:spec3D}(b) it is clear that the collective noise (due to atoms and photons) results in a suppression of photon-flux, consistent with observations of recent experiment~\cite{tilman_2013}. We note that the suppression of flux observed here is specific to the parameter values used and may turn into enhancement in other regimes.

\textit{Conclusion -}
In this paper, we studied the critical dynamics of density fluctuations of a driven condensate in an optical cavity by taking into account the full spectrum of collective motional modes of the condensate. We show that the cavity induces long-range two-body interactions between the atoms of a condensate and that the interaction is proportional to the photon Green's function of the cavity. 
In the adiabatic limit where the cavity field is eliminated, this leads to a softening of a roton-like collective mode recently observed in experiments. Going beyond the two-mode (Dicke limit) description and including the full spectrum of atomic modes, we show that the roton-like mode of the system couples to other atomic modes in a large but finite system and hence acquires a finite lifetime that non-trivially depends on the pump strength. We find that the collisional s-wave interaction is a relevant perturbation which determines the nature of critical dynamics. Because both finite size and interactions are relevant to recent experiments, we believe that an ab-initio description of the atomic dynamics is crucial. Indeed, our results shed light on the strongly pump dependent atomic damping observed in recent experiments \cite{tilman_2013} that lead to the suppression of sub-threshold cavity field fluctuations below what's expected from cavity leakage alone \footnote{2D extension of this work shows that features depending on unitary dynamics remain dimensionally invariant. However, features arising due to non-unitary dynamics such as the exponent associated with vanishing damping have dimensional dependence. This work will be presented elsewhere.}.

\section{Acknowledgements}

We thank David Huse, Marco Schiro and Emanuele Dalla Torre for useful discussions.  This work was supported by the US National Science Foundation through the NSF CAREER Grant No. DMR-1151810 and by the Swiss NSF through Grant No. PP00P2-123519/1.

\bibliographystyle{apsrev}
\bibliography{bec_cavity}

\end{document}